\documentclass[10pt,conference]{IEEEtran}
\IEEEoverridecommandlockouts
\usepackage{cite}
\usepackage{amsmath,amssymb,amsfonts}
\usepackage{algorithmic}
\usepackage{graphicx}
\usepackage{textcomp}
\usepackage{xcolor}
\usepackage{caption}

\usepackage{epsfig}
\usepackage{graphicx}        
\usepackage{multicol}        
\usepackage[bottom]{footmisc}
\usepackage{url}
\usepackage{comment}
\usepackage{multirow, booktabs}
\usepackage{amsmath}
\usepackage{amssymb}
\usepackage{color}
\usepackage{flushend}
\usepackage{enumerate}
\usepackage{bm}
\usepackage{here}
 \usepackage{ascmac}
 \usepackage{fancybox}
 \usepackage{subfigure}
\usepackage{hyperref}
\usepackage{makecell, color, colortbl}

\newcommand{\firstRQ}{How often do patches receive divergent scores?}
\newcommand{\secondRQ}{How often are patches with divergent scores eventually integrated?}
\newcommand{\thirdRQ}{How are reviewers involved in patches with divergent scores?}
\newcommand{\fourthRQ}{What drives patches with divergent scores to be abandoned?}
\newcommand{\fifthRQ}{What concerns are resolved in patches with divergent scores that are eventually integrated?}

\newcommand{\os}{\textsc{OpenStack}}
\newcommand{\nv}{\textsc{Nova}}

\newcommand{\qt}{\textsc{Qt}}
\newcommand{\qb}{\textsc{QtBase}}

\def\BibTeX{{\rm B\kern-.05em{\sc i\kern-.025em b}\kern-.08em
    T\kern-.1667em\lower.7ex\hbox{E}\kern-.125emX}}
\begin{document}

\title{Code Reviews with Divergent Review Scores:\\ \LARGE{An Empirical Study of the OpenStack and Qt Communities}}

\author{%
    \IEEEauthorblockN{Toshiki Hirao\IEEEauthorrefmark{1}, Shane McIntosh\IEEEauthorrefmark{2}, Akinori Ihara\IEEEauthorrefmark{3}, Kenichi Matsumoto\IEEEauthorrefmark{4}}\\
    
    \IEEEauthorblockA{%
        \IEEEauthorrefmark{1}dTosh Inc., Japan; toshiki.hirao@dtosh.com\\
        \IEEEauthorrefmark{2}University of Waterloo, Canada; shane.mcintosh@uwaterloo.ca\\
        \IEEEauthorrefmark{3}Wakayama University, Japan; ihara@wakayama-u.ac.jp\\
        \IEEEauthorrefmark{4}Nara Institute of Science and Technology, Japan; matumoto@is.naist.jp
    }
}

\maketitle



\IEEEraisesectionheading{\section{Abstract}\label{Abstract}}
Code review is widely considered a best practice for software quality assurance~\cite{Wiegers_AWLP2002}.
The Modern Code Review (MCR) process---a lightweight variant of the traditional code inspection process~\cite{Fagan_IBM1976}---allows developers to post patches for review.
Reviewers (i.e., other team members) are either:
(1) appointed automatically based on their expertise~\cite{Thongtanunam_SANER2015, Zanjani_TSE2015, Chanchal_ICSE2016};
(2) invited by the author~\cite{Thongtanunam_SANER2015, McIntosh_MSR2014, Hirao_OSS2016};
or (3) self-selected by broadcasting a review request to a mailing list~\cite{Rigby_ICSE2008, Rigby_ICSE2011, Guzzi_MSR2013}.

Reviewer opinions about a patch may differ.
Divergent reviews can slow integration processes down~\cite{Rigby_ICSE2011} and can create a tense environment for contributors~\cite{Sadowski_icse2018}.
For instance, consider review \#12807 from the \qb\ project.\footnote{\label{id_12807}\url{https://codereview.qt-project.org/\#/c/12807/}}
The first reviewer approves the patch for integration (+2).
Afterwards, another reviewer blocks the patch from being integrated with a strong disapproval (-2), arguing that the scope of the patch must be expanded before integration could be permitted.
Those reviewers who provided divergent scores discussed whether the scope of the patch was sufficient for five days, but an agreement was never reached.
One month later, the patch author abandoned the patch without participating in the discussion.
Despite making several prior contributions, this is the last patch that the author submitted to the \qb\ project.

We set out to better understand patches with divergent review scores and the process by which integration decisions are made.
To do so, we analyze the large and thriving \os\ and \qt\ communities.
Through quantitative analysis of 49,694 reviews, we address the following research questions:

\begin{enumerate}[{\bf (RQ1)}]
\item{\bf \firstRQ} \\
 \underline{Motivation:}
   Review discussions may diverge among reviewers.
   We first set out to investigate how often patches with divergent review scores occur.

   \noindent\underline{Results:}
   Divergent review scores are not rare. Indeed, 15\%--37\% of the studied patch revisions that receive review scores of opposing polarity.

 \item{\bf \secondRQ}\\
   \underline{Motivation:}
   Given that patches with divergent scores receive both positive and negative scores, making an integration decision is not straightforward.
   Indeed, integration decisions do not always follow a simple majority rule~\cite{Hirao_CASCON2015}.
   We want to know how often these patches are eventually integrated.

   \noindent\underline{Results:}
   Patches are integrated more often than they are abandoned.
   For example, patches that elicit positive and negative scores of equal strength are eventually integrated on average 71\% of the time.
    The order in which review scores appear correlates with the integration rate, which tends to increase if negative scores precede positive ones.

   \item{\bf \thirdRQ} \\
    \underline{Motivation:}
    Patches may require scores from additional reviewers to arrive at a final decision, imposing an overhead on development.
    In reviews with divergent scores, we set out to study (a) if additional reviewers are involved; (b) when reviewers join the reviews; and (c) when divergence tends to occur.

    \noindent\underline{Results:}
    Patches that are eventually integrated involve one or two more reviewers than patches without divergent scores on average.
    Moreover, positive scores appear before negative scores in 70\% of patches with divergent scores.
    Reviewers may feel pressured to critique such patches before integration (e.g., due to lazy consensus).\footnote{\url{https://community.apache.org/committers/lazyConsensus.html}}
    Finally, divergence tends to arise early, with 75\% of them occurring by the third (\qt) or fourth (\os) revision.
\end{enumerate}

To better understand divergent review discussions, we qualitatively analyze: (a) all 305 of the patches that elicit strongly divergent scores from members of the core development teams; (b) a random sample of 630 patches that elicit weakly divergent scores from contributors; and (c) a random sample of 305 patches without divergent scores.
In doing so, we address the following research questions:

\begin{enumerate}[{\bf (RQ1)}]
   \setcounter{enumi}{3}
 \item {\bf \fourthRQ}\\
 \underline{Motivation:}
 In RQ2, we observe that 29\% of the studied patches with divergent scores are eventually abandoned.
 Since each patch requires effort to produce, we want to understand how the decision to abandon patches with divergent scores is reached.

 \noindent\underline{Results:}
 Abandoned patches with strong divergent scores more often suffer from {\em external} issues than patches with weakly divergent scores and without divergent scores do.
 These external issues most often relate to release planning and the concurrent development of solutions to the same problem.

\item {\bf \fifthRQ}\\
 \underline{Motivation:}
 In the 71\% of patches with divergent scores that are eventually integrated (see RQ2), the reviewer concerns are being addressed.
 We set out to study which types of concerns are typically addressed.
 
 \noindent\underline{Results:}
 In \os\ and \nv, reviewer concerns are more often indirectly addressed (e.g., through integration timing) in patches with strong divergent scores than patches with weakly divergent and without divergent scores.
 On the other hand, in \qb, reviewer concerns are often directly addressed through patch revision, irrespective of whether divergent scores are present.
\end{enumerate}

  Based on our results, we suggest that:
  (a) software organizations should be aware of the potential for divergent review discussion, since patches with divergent scores are not rare and tend to require additional personnel to be resolved;
  (b) automation could relieve the burden of reviewing for external concerns;
  and (c) authors should note that even the most divisive patches are often integrated through constructive discussion, integration timing, and careful revision.

\begin{table}[t]
    \centering
    \fontsize{10}{11}\selectfont
    \captionsetup{font=large}
    \caption*{{\bf{Satisfying the criteria for journal first presentation at ICSE2021.}}}
    \label{tab:criteria}
   
    \begin{tabular}{ p{4.1cm} | p{4.1cm}} \hline
    {\bf{Criterion}} & {\bf{Response}} \\ \hline
    The associated accepted journal paper was accepted to a journal from the list below no earlier than November 1st, 2019 and no later than November 1st, 2020: & The article was accepted for publication in the IEEE Transactions on Software Engineering on February 25th, 2020. \\ \hline
    The paper is in the scope of the conference. & The paper fits under the following topics of interest that appear in the ICSE 2021 call for papers: ``Mining software repositories'' and ``Evolution and maintenance''. \\ \hline
    The paper reports completely new research results and/or presents novel contributions that significantly extend and were not previously reported in prior work. & Yes, this paper makes several novel contributions (see Section 1 of the paper for an overview). \\ \hline
    The paper does not extend prior work solely with additional proofs or algorithms (or other such details presented for completeness), additional empirical results, or minor enhancements or variants of the results presented in the prior work. & This paper is not an extended version of our prior work or that of others. \\ \hline
    The paper has not been presented at, and is not under consideration for, journal-first programs of other conferences. & ICSE 2021 would be the first conference at which this work would be presented.\\ \hline
    The paper should not exclusively report a secondary study, e.g., systematic reviews, mapping studies, surveys. & As far as we know, this paper is not an exclusion from any of other studies. \\  \hline
    \end{tabular}
\end{table}

\label{intro}








\bibliographystyle{IEEEtran}
\bibliography{IEEEabrv,./bibfile.bib}

\end{document}